\documentclass[10pt,preprint2]{aastex}
\usepackage{epsfig}           

\shorttitle{On the large-scale angular distribution of SGRB}
\shortauthors{Bernui et al.}

\begin{document}

\title{On the large-scale angular distribution of short-Gamma ray bursts}

\author{A. Bernui, I. S. Ferreira, and C. A. Wuensche}

\affil{Instituto Nacional de Pesquisas Espaciais, 
Divis\~ao de Astrof\'{\i}sica, \\
Av. dos Astronautas 1758, CEP 12227-010
S\~ao Jos\'e dos Campos, SP, Brazil}
\email{bernui@das.inpe.br; ivan@das.inpe.br; alex@das.inpe.br}

\begin{abstract}
We investigate the large-scale angular distribution of the short-Gamma ray 
bursts (SGRBs) from BATSE experiment, using a new coordinates-free method. 
The analyses performed take into account the angular correlations induced 
by the non-uniform sky exposure during the experiment,  
and the uncertainty in the measured angular coordinates. 
Comparising the large-scale angular correlations from the data with those 
expected from simulations using the exposure function we find similar 
features. 
Additionally, confronting the large-angle correlations computed from 
the data with those obtained from simulated maps produced under the assumption 
of statistical isotropy we found that they are incompatible at 95\% confidence 
level.
However, such dif\/ferences are restricted to the angular scales 
$36^{\circ} - 45^{\circ}$, which are likely to be due to the non-uniform sky 
exposure. 
This result strongly suggests that the set of SGRBs from BATSE are intrinsically 
isotropic. 
Moreover, we also investigated a possible large-angle correlation of these data 
with the supergalactic plane. 
No evidence for such large-scale anisotropy was found. 
\end{abstract}

\section{Introduction}  \label{Introduction}

The apparent isotropy in the large-scale angular distribution of the Gamma 
ray bursts (GRBs) is a long-standing debate~\citep{Meegan92,Briggs96,Tegmark,
Piran-Singh,Metzger,Balazs,Meszaros}. 
Since the first detection with the VELA satellite~\citep{VELA} the origin of 
these highly energetic events has remained a challenge. 
Even if the origin of GRBs turns out to be extra-galactic or cosmological, 
as suggested by current data (see, e.g.,~\citet{Piran,review1,review2,review3} 
and references therein), this does not ensure that their distribution is 
isotropic. 
Up to now, no dominant anisotropies has been found in the angular distribution 
of GRBs. 
However, if detected, small anisotropic effects may reveal valuable information 
about their origin. 
Additionally, the discovery of a large angular scale pattern in the sky 
distribution of GRBs may be useful to identify their sources by 
cross-correlating them with catalogs of cosmic objects, e.g., early-type 
galaxies, hard X-ray sources, etc.~\citep{Briggs96,Tegmark,Piran,review3}. 

The reported statistical analyses of the all-sky survey data from BATSE show
that their large-scale angular distribution is consistent with 
isotropy~\citep{Piran,review3}, although aspects like observational artifacts 
have not been fully considered in some of these studies. 
It is well known that anisotropies with distinct origins manifest themselves 
on different angular scales and with different magnitudes. 
In this connection, it is reasonable to consider different approaches that 
can, in principle, provide information about multiple types of anisotropy, 
imprinted as angular correlation signatures (ACS), that may be possibly 
present in the data. 

Here we apply a new coordinates-free method to search for large-scale ACS 
($\geq 18^{\circ}$) in a subset of the BATSE GRB data~\citep{Meegan00}, 
namely the Short-GRBs, and then investigate their significance levels 
through the comparison with a large number of Monte Carlo maps. 
Such simulated maps were produced under similar conditions as the catalog 
under analysis, that is, taking into account the non-uniform sky exposure 
of BATSE and the uncertainty in the coordinates measurements.
Furthermore, for completeness, we also compare the ACS of the catalog of 
GRBs with those corresponding to statistically isotropic Monte Carlo maps. 
Finally, we also investigated the possible large-scale angular correlation 
between the set of Short-GRBs and the supergalactic plane, in an attempt to 
search for likely host galaxies of these events (as recently suggested 
by~\citet{Ghirlanda}). 

The outline of this paper is the following: 
in section~\ref{BATSE-catalog} we use GRBs data from the BATSE experiment to 
determine the Short-GRBs catalog to be investigated, and in section~\ref{method}, 
we describe the method employed in such studies.
The data analyses and results are shown in section~\ref{analyses}, 
and finally in section~\ref{conclusions} we formulate our conclusions.

\section{The Short-GRBs from BATSE catalog}  \label{BATSE-catalog}

The physical analysis of GRBs utilizes their temporal and spectral 
properties~\citep{Fishman-Meegan}. 
Despite the dif\/ferent light-curves observed in the spectra of GRBs, 
a useful parameter to classify them is the burst duration $T_{90}$, defined 
as the time interval during which 90\% of the f\/luence is measured.
The current BATSE catalog 
4Bc $\!$\footnote{http://gammaray.msfc.nasa.gov/batse/grb/catalog/current/} 
contains 2\,702 events, where only 2\,037 GRBs have their parameter $T_{90}$ 
measured~\citep{Meegan00}. 

At first, the $T_{90}$ value was used to divide the set of GRBs into two 
dif\/ferent sub-classes: 
the Short-GRBs (SGRBs), with $T_{90} < 2$s, and the Long-GRBs, 
with $T_{90} \gtrsim 10$s~\citep{Kouveliotou,review1,review2,review3}. 
However, the use of this definition of SGRBs is instrument dependent and is 
susceptible to observational biases~\citep{Hakkila07b}.
For this reason, one should consider in addition to the $T_{90}$ criterium the 
parameter $HR_{3/21}$~\citep{Mukherjee} which is defined as the 100 to 300 keV 
f\/luence divided by the 25 to 100 keV f\/luence of each GRB in 
BATSE 4Bc~\citep{Hakkila07a,Hakkila07b}. 
Thus, the appropriate definition for a catalog of SGRBs is~\citep{Hakkila07a,Hakkila07b}: 
${\cal C}=$ \{516 events with $2$s $\leq T_{90} < 4.7$s and $HR_{3/21} > 3$\}. 

With this information, and using a new coordinates-free method to be described 
in the next section, we shall perform a detailed analysis of the large-scale 
ACS present in the sky distribution of the SGRBs from BATSE.

\section{The 2PACF and the Sigma-Map analysis}  \label{method}

Let $\Omega_j^{\gamma_0} \equiv \Omega(\theta_j,\phi_j;\gamma_0) \in
{\cal S}^2$ be a spherical cap region on the celestial sphere, 
of $\gamma_0$ degrees of aperture, with vertex at the $j$-th pixel, 
$j=1, \ldots, N_{\mbox{\footnotesize caps}}$, where $(\theta_j,\phi_j)$ 
are the angular coordinates of the center of the $j$-th pixel.
Both, the number of spherical caps $N_{\mbox{\footnotesize caps}}$ and the 
coordinates of their center $(\theta_j,\phi_j)$ are defined using the HEALPix 
pixelization scheme~\citep{Gorski}. 
The spherical caps are such that their union completely covers the celestial 
sphere ${\cal S}^2$. 

Let ${\cal{C}}^{\,j}$ be the catalog of cosmic objects located in the $j$-th 
spherical cap $\Omega_j^{\gamma_0}$. 
The 2PACF of these objects~\citep{Chen-Hakkila,Padmanabhan}, denoted as 
$\Upsilon_j(\gamma_i;\gamma_0)$, is the difference between the normalized 
frequency distribution and that expected from the number of pairs-of-objects 
with angular distances in the interval 
$(\gamma_i - 0.5\delta,\gamma_i + 0.5\delta],\, i=1,\ldots,N_{\mbox{\footnotesize bins}}$, 
where $\gamma_i \equiv (i-0.5)\delta$ and 
$\delta \equiv 2\gamma_0 / N_{\mbox{\footnotesize bins}}$ 
is the bin-width. 
The expected frequency distribution is achieved by a large number of Monte 
Carlo realizations of isotropically distributed objects in $\Omega_j^{\gamma_0}$, 
containing a similar number of objects as in ${\cal{C}}^{\,j}$~\citep{Teixeira,BV}. 
The 2PACF has the property that its mean, obtained by integrating over all 
separation angles~\citep{Chen-Hakkila}, is zero. 
A positive (negative) value of $\Upsilon_j$ indicates that objects with these 
angular separations are correlated (anti-correlated), while zero indicates 
no correlation.

Define now the scalar function $\sigma: \Omega_j \mapsto {\Re}^{+}$, for
$j=1, \ldots, N_{\mbox{\footnotesize caps}}$, which assigns to the 
$j$-cap, centered at $(\theta_j,\phi_j)$, a real positive number 
$\sigma_j \equiv \sigma(\theta_j,\phi_j) \in \Re^+$. 
The most natural way of defining a measure $\sigma$ is through the 
variance of the $\Upsilon_j$ function, we thus define~\citep{BMRT} 
\begin{equation} \label{sigma}
\sigma^2_j  \equiv \frac{1}{N_{\mbox{\footnotesize bins}}}
\sum_{i=1}^{N_{\mbox{\footnotesize bins}}} \Upsilon^2_j (\gamma_i;\gamma_0).
\end{equation}
To obtain a quantitative measure of the ACS of the GRBs sky map, we cover 
the celestial sphere with $N_{\mbox{\footnotesize caps}}$ spherical caps, 
and calculate the set of values $\{ \sigma_j, \, j=1,...,N_{\mbox{\footnotesize caps}} \}$
using eq.~(\ref{sigma}).
Patching together the set $\{ \sigma_j \}$ in the celestial sphere 
according to a coloured scale (where, for instance, 
$\sigma^{\mbox{\footnotesize minimum}} \rightarrow blue$, 
$\sigma^{\mbox{\footnotesize maximum}} \rightarrow red$) we obtain a sigma-map. 
Finally, we quantify the ACS of a given sigma-map by calculating its angular 
power spectrum. 
Because the sigma-map assigns a real value to each pixel in the celestial 
sphere, that is $\sigma = \sigma(\theta,\phi)$, one can expand it in 
spherical harmonics: 
$\sigma(\theta,\phi) = \sum_{\ell,\, m} A_{\ell\, m} Y_{\ell\, m}(\theta,\phi)$ 
where the set of values $\{ S_{\ell},\, \ell=0,1,2,... \}$, defined by 
$S_{\ell} \equiv (1 / (2\ell+1)) \sum_{m={\mbox{\small -}}\ell}^{\ell} \, |A_{\ell\, m}|^2$, 
is the angular power spectrum of the sigma-map. 
Because we are interested in the large-scale angular correlations, we shall 
concentrate on $\{ S_{\ell}, \ell = 1,2,...,10 \}$.
Notice that we are interested in the angular power spectrum of the sigma-map, 
and not that of the celestial sphere where the GRB events are located; 
this later case was already done by \citet{Briggs96} and \citet{Tegmark}. 
As we shall see, the sigma-map analysis is able to reveal very small 
anisotropies, like those induced by the BATSE's sky exposure, despite the 
small burst detection rate of the SGRBs.

\section{Data analyses and results}  \label{analyses}

In this section we shall apply the sigma-map method explained in the previous 
section to study the large-scale ACS present in the angular distribution of the 
516 SGRBs listed in the catalog ${\cal C}$. 
In the following, all the sigma-maps were calculated using spherical caps of 
$\gamma_0 = 90^{\circ}$ of aperture, that is hemispheres 
(smaller spherical caps have less SGRBs in each 
${\cal{C}}^{\,j}, \, j=1,\ldots,N_{\mbox{\footnotesize caps}}$, hence 
produce large statistical noise in the $\Upsilon_j$ functions). 
We also used $N_{\mbox{\footnotesize bins}}=90$ and 
$N_{\mbox{\footnotesize caps}}=768$ in these analyses. 

An important issue that deserves close inspection is the presence of 
anisotropic ACS in the data induced by the non-uniform sky exposure (NUSE) 
during the BATSE experiment~\citep{Hakkila98}, expected because some latitudes 
of the sky were over-observed while others were under-observed. 
Because there are no reports quantifying or tracing out the influence of the NUSE 
at large angular scales in the current BATSE catalog 4Bc (see~\citet{Chen-Hakkila} 
for analyses of the 3B and 4B catalogs) it is interesting to use the sigma-map 
method to investigate the possible anisotropic angular correlations that may be 
present in the data even if their magnitudes are small. 
For this, our strategy to reveal the large-scale ACS in the data runs in three 
steps.
First, we produce $1\,000$ Monte Carlo maps simulating the sky positions of 
516 cosmic objects according to the NUSE function~\citep{Hakkila98}, then we 
calculate in each case the corresponding sigma-map, and finally we compute the 
angular power spectrum $\{ S_{\ell}, \ell=1,...,10 \}$ of each of these 
sigma-maps. 
Second, we generate $1\,000$ Monte Carlo maps simulating the sky positions of 
516 isotropically distributed cosmic objects, then we compute for each of these 
Monte Carlo maps their corresponding sigma-maps, and finally we calculate the 
angular power spectrum of these sigma-maps. 
Third, we calculate the sigma-maps, and their respective angular power spectrum, 
of the SGRBs listed in catalog ${\cal C}$.

\begin{figure}   
\plotone{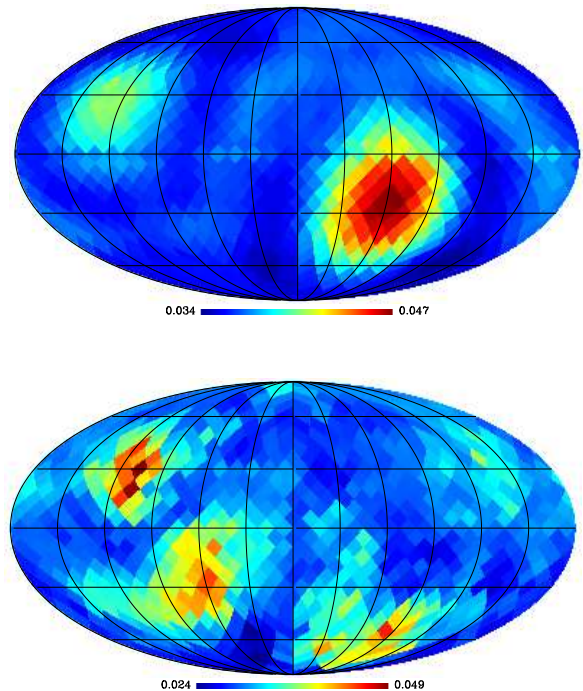}
\caption{
Sigma-maps in galactic coordinates and with a graticule of $30^{\circ}$. 
Top: This map is the average of 100 of sigma-maps each computed from a Monte 
Carlo simulation of the sky distribution of SGRB events according to the NUSE 
function.
Bottom: This is the sigma-map calculated from the catalog ${\cal C}$ of 
BATSE SGRBs data.
}
\label{figure1}
\end{figure}

In figure~\ref{figure1} we show two sigma-maps in galactic coordinates.
In the top panel, we show the average of $100$ sigma-maps, randomly 
chosen in between the $1\,000$ sigma-maps computed from a similar number of 
Monte Carlo sky maps which simulate different catalogs of SGRB according to 
the NUSE function. 
In the bottom panel we exhibit the sigma-map corresponding to the catalog ${\cal C}$. 

\begin{figure}  
\plotone{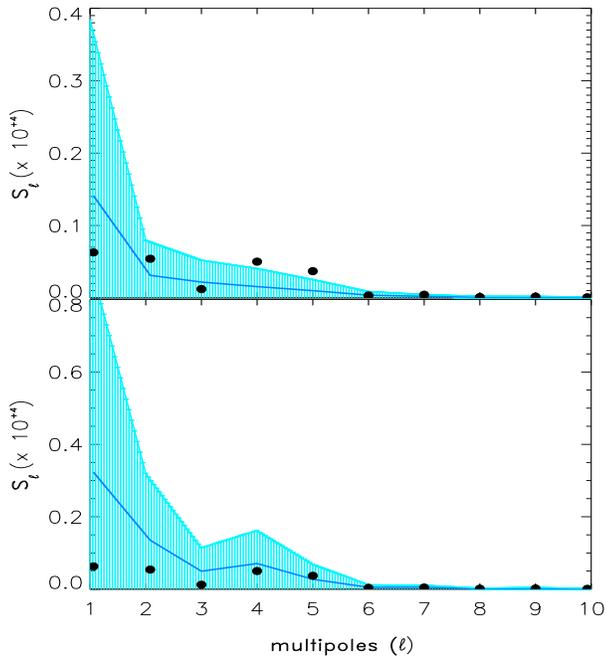}
\caption{Averaged angular power spectra of $1\,000$ sigma-maps computed from 
equal number of Monte Carlo maps, which were produced considering two cases: 
isotropic (top) and anisotropic (bottom) hypotheses. 
For the anisotropic case, the Monte Carlo maps were generated using the 
NUSE function. 
Together we plot the angular power spectrum of the sigma-map, represented 
as bullets, corresponding to the BATSE SGRBs listed in the catalog ${\cal C}$. 
The shadowed regions correspond to 2 standard deviations in each case. 
}
\label{figure2}
\end{figure}

In figure~\ref{figure2} we display a comparative analyses, taking into account 
isotropic and non-isotropic cases, of the angular power spectrum of the sigma-map 
obtained from the angular distribution of the SGRBs listed in ${\cal C}$. 
In the top panel, we plotted the angular power spectrum $S_{\ell}$ versus 
$\ell$ of the sigma-map obtained from the catalog ${\cal C}$, together with 
the mean of $1\,000$ sigma-maps computed from an equal number of statistically 
isotropic Monte Carlo sky maps. 
In the bottom panel, the plot is similar except that the Monte Carlo sky maps 
have anisotropic ACS because were produced considering the NUSE function 
of BATSE. 
In both plots the shadowed areas correspond to 2 standard deviations.
Besides some small dif\/ferences, the angular power spectra corresponding to 
the sigma-maps computed from the SGRBs show a very similar large-scale 
structure when compared with the mean angular power spectrum of the 
sigma-maps obtained from Monte Carlos produced according to the NUSE function. 

A comparative analysis of the ACS corresponding to these cases, isotropic 
and non-isotropic due to NUSE function, is better seen if we plot 
$\ell \, (\ell+1) \, S_{\ell}$ versus $\ell$, as showed in 
figure~\ref{figure3}. 
There we display the correponding data from the SGRBs together 
with the mean of the angular power spectra of the isotropic and non-isotropic 
cases, where now the shadowed area corresponds to 2 standard deviations of the 
isotropic case. 
As observed, the data have a very similar behavior to the non-isotropic 
case, and is different from the isotropic case which shows a f\/lat spectrum.
Thus, data and simulated isotropic maps are incompatible at 95\% confidence 
level. 
However such dif\/ferences are mainly restricted to the angular scales 
$36^{\circ} - 45^{\circ}$ which exactly reproduce the imprints exhibited by 
the angular power spectrum of the non-isotropic case. 
In the absence of ACS other than those expected by the NUSE of the BATSE 
experiment, this result strongly suggests that the SGRBs are intrinsically 
isotropic. 

To test the robustness of our calculations we also performed the sigma-map analyses 
with $N_{\mbox{\footnotesize bins}}=180$ and $N_{\mbox{\footnotesize caps}}=3\,072$, 
obtaining the same result. 

Furthermore, we also searched for a possible correlation between the SGRBs listed 
in ${\cal C}$ with the supergalactic plane, where nearby galaxies appear to be more
concentrated. 
Because we do not know how many events could be originated in these galaxies, 
we generate three sets of 300 Monte Carlos considering in each case a different 
number of simulated GRBs provided by an anisotropic distribution which selects 
events near the supergalactic plane.
That is, we generated sets of maps where 33\%, 50\%, and 66\% of the events 
were produced by such anisotropic distribution, respectively.  
We then computed their corresponding sigma-maps in order to compare them with 
the sigma-map calculated from the SGRBs listed in ${\cal C}$.
To measure such a possible correlation we computed the linear Pearson correlation 
coefficient between the sigma-map of the SGRBs and each one of the sigma-maps 
obtained from these sets of Monte Carlo realizations. 
Notice that such a coefficient varies from $0$ (for totally uncorrelated maps) 
to $1$ (for fully correlated maps). 
Our results show that the Pearson's coefficient is, in mean, less than $0.03$ 
(using 300 Monte Carlos for each of the three above mentioned 
cases).  
To realize whether this value is statistically significant, we computed the 
Pearson's coefficient in some illustrative cases. 
For example, the mean Pearson's coefficient correlating one sigma-map, coming 
from a given set of sigma-maps computed from the above mentioned 66\% anisotropic 
Monte Carlo maps, with the rest of sigma-maps from such set is $0.23 \pm 0.16$ 
(the result is similar in the other two cases). 
On the other hand, the mean Pearson's coefficient correlating one sigma-map, 
chosen randomly from the set of $1\,000$ sigma-maps calculated from Monte 
Carlo maps produced according to the NUSE function, with the rest of sigma-maps 
of this set is $0.25 \pm 0.16$. 
Similarly, the mean Pearson's coefficient correlating any sigma-map, from the 
set of $1\,000$ sigma-maps computed from Monte Carlo statistically isotropic 
maps, with the rest of sigma-maps of this set is $0.15 \pm 0.12$. 
 
Moreover, comparing the sigma-map computed from BATSE SGRBs catalog ${\cal C}$ 
with each of the $1\,000$ sigma-maps, obtained from Monte Carlo maps generated 
according to the NUSE function, the mean Pearson's coefficient is $0.14 \pm 0.11$.
A similar analysis of the sigma-map of BATSE SGRBs but now considering those 
Monte Carlo maps generated under the statistical isotropy hypothesis, tells 
us that the Pearson's coefficient is, in mean, $0.12 \pm 0.09$. 
Taken together this information suggests that there is no evidence for a 
large-scale angular correlation between BATSE SGRBs and simulated maps 
produced considering different amounts of events coming from an anisotropic 
distribution that selects positions in the supergalactic plane and its surrounds.
Notice that this result does not contradict the correlation found 
by~\citet{Ghirlanda} which is valid for small angular distances $\leq 3^{\circ}$, 
while the present analysis concerns angular scales $\geq 18^{\circ}$. 

Finally, we also tested the robustness of our results under the change of the 
angular coordinates of the BATSE SGRBs due to the measured error 
boxes~\citep{Briggs98}. 
This was done by sorting their angular coordinates within the limits given by 
the error boxes ($\pm 2.5^{\circ}$) we found no measurable dif\/ference with 
respect to the results presented here. 

\section{Conclusions}    \label{conclusions}

The purpose of this study is to know the large-scale angular correlations of the 
set of 516 BATSE SGRBs, and to discover if these correlations are compatible with 
a statistically isotropic distribution of events, or instead they reveal the ACS 
resulting as a consequence of the NUSE function of BATSE experiment. 
To elucidate this, we need to know the angular power spectra of two sets of 
sigma-maps: 
one set is computed from statistically isotropic Monte Carlo maps and the other 
is calculated from Monte Carlo maps that simulate the sky position of the events 
using the NUSE function of BATSE experiment. 
After that, we compare the power spectra of these sigma-maps with the angular 
power spectrum of the sigma-map computed from the BATSE SGRBs data. 

\begin{figure} 
\plotone{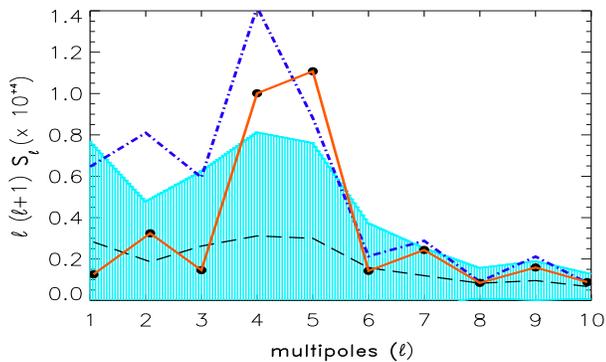}
\caption{Plot of the angular power spectra in the form 
$\ell \, (\ell+1) \, S_{\ell}$ versus $\ell$.
The dashed (dot-dashed) line corresponds to the mean angular power spectrum 
of $1\,000$ sigma-maps, each one computed from a Monte Carlo isotropic map 
(anisotropic map, according to the NUSE function). 
The shadowed region corresponds to 2 standard deviations of the isotropic 
case. 
Together with these data we plotted the angular power spectrum of the 
SGRBs listed in ${\cal C}$, represented as bullets. 
}
\label{figure3}
\end{figure}

The first thing to be noticed in the angular power spectrum of the BATSE SGRBs, 
plotted in figure~\ref{figure2}, is the absence of dominant anisotropies at the 
largest angular scales $60^{\circ} - 180^{\circ}$ ($\ell = 1, 2, 3$), and a 
similar situation for angular scales $\leq 30^{\circ}$ ($\ell \geq 6$). 
However, peculiar features appear at the angular scales $36^{\circ} - 45^{\circ}$  
($\ell = 4, 5$), and for a better understanding of what information is encoded 
there we plot these data in the form 
$\ell \, (\ell+1) \, S_{\ell}$ versus $\ell$ (see figure~\ref{figure3}). 
In fact, figure~\ref{figure3} reveals a second interesting thing, that is, 
the non-flat spectrum of the BATSE data (represented by bullets) which clearly 
dif\/fers from the flat angular power spectra showed by the statistically 
isotropic Monte Carlo data (the dashed line). 
We also observe in figure~\ref{figure3} that the mean of the angular power 
spectra of the sigma-maps computed from Monte Carlo maps generated according 
to the NUSE function (the dot-dashed line) has also a non-flat spectrum which 
is very similar to the corresponding one obtained from BATSE SGRBs. 
In other words, the large-scale angular correlations of the BATSE SGRBs exhibit 
the anisotropic imprints expected in the data due to the NUSE of BATSE experiment. 
Other ACS are not found to be statistically significant, at 95\% CL. 
In conclusion, these results strongly suggests that the SGRBs are intrinsically 
isotropic. 

Finally, we also studied the possible large-angle correlation between 
the SGRBs data and Monte Carlos with (different amounts of) simulated events 
concentrated towards the supergalactic plane. 
No evidence for such large-scale anisotropy was found in the BATSE SGRBs.

\section*{Acknowledgments}
\noindent
We acknowledge use of the BATSE data~\citep{Meegan00}. 
Some of the results in this paper have been derived using the HEALPix 
package~\citep{Gorski}.
We thank T. Villela, Z. Bagoly, A. Teixeira, R. Tavakol, E. Berger, 
and B. Schaefer for insightful comments and suggestions. 
We also acknowledge Prof. J. Hakkila for very healpful exchanges regarding 
BATSE data. 
We are indebted to the anonymous referee for valuable suggestions and 
constructive criticisms. 
CAW was partially supported by CNPq grant 307433/2004-8.
ISF and AB acknowledge CAPES and PCI/DTI-MCT fellowships, respectively.

\end{document}